\documentstyle[aps,preprint]{revtex}

\newcommand{\ba}{\begin{eqnarray}}
\newcommand{\be}{\begin{equation}}
\newcommand{\ea}{\end{eqnarray}}
\newcommand{\ee}{\end{equation}}
\newcommand{\er}{\end{eqnarray}}
\newcommand{\br}{\begin{eqnarray}}
\newcommand{\dslash}{\partial\!\!\!/}
\newcommand{\aslash}{A\!\!\!/}

\title{{\bf Interference Phenomenon for the Faddeevian Regularization of 2D Chiral Fermionic Determinants}}
\author{Everton M. C. Abreu $^a$\footnote{Financialy supported by Funda\c{c}\~ao de Amparo \`a Pesquisa do Estado de S\~ao Paulo (FAPESP), Grant No. 99/03404-6.}, 
Anderson Ilha${}^{b}$, Clifford Neves${}^{b}$  
and Clovis Wotzasek${}^{b}$\thanks{respectively: everton@feg.unesp.br;\, ais@if.ufrj.br; \,clifford@if.ufrj.br; \,clovis@if.ufrj.br}}
\address{ ${}^{a}$ Departamento de F\'\i sica e Qu\'\i mica, Universidade Estadual Paulista,\\
 Av. Ariberto Pereira da Cunha 333, Guaratinguet\'a, 12500-000, \\ S\~ao Paulo, SP, Brazil \\
${}^{b}$ Instituto de F\'{\i}sica, Universidade Federal do Rio de Janeiro, Caixa Postal 68528, 21945-970, Rio de Janeiro, RJ, Brazil }

\date{\today}

\begin{document}

\maketitle

\begin{abstract}
The classification of the regularization ambiguity of 2D fermionic determinant in three
different classes according to the number of second-class constraints, including the new faddeevian regularization, is examined and extended.
We found a new and important result that the faddeevian class, with three second-class
constraints, possess a free continuous one parameter family of 
elements. The criterion of unitarity restricts the parameter to the same range found earlier by Jackiw and Rajaraman for the two-constraints class.
We studied the restriction imposed by the
interference of right-left modes of the chiral Schwinger model ($\chi QED_{2}$) using Stone's soldering formalism. 
The interference effects between 
right and left movers, producing the massive vectorial photon, are shown to constrain the regularization parameter to belong to the four-constraints 
class which is the only non-ambiguous class with a unique regularization parameter.
\end{abstract}
\bigskip

PACS:{11.10.Ef,11.15.-q,11.30.Rd,11.40.Ex}

\newpage

\bigskip


\section{Introduction}

It is often claimed that chiral interaction of two-dimensional fermionic gauge models poses
an obstruction to gauge symmetry.  In this paper we clarify several aspects of this question
for different regularizations of the chiral fermionic determinant, including the new Faddeevian
regularization case proposed by Mitra\cite{PM}, under the point of view of the Stone's soldering formalism\cite{ms}.
It is worth mentioning that understanding the properties of 2D fermionic actions is crucial
in several aspects. For instance, the 1-cocycle necessary in recent
discussions on smooth functional bosonization \cite{DNS,DS}, which is just the expression of the 2D
anomaly, is known to be the origin of higher dimensional anomalies through a set of descent equations\cite{DTMP}.
Incidentally, the anomaly phenomenon still defies a complete explanation.

This paper is devoted to analyze and explore the 
restrictions that the soldering mechanism \cite{ms,ADW,W,AW} imposes 
over the regularization ambiguity of 2D chiral fermionic determinants.
The soldering technique that is dimensionally independent and designed 
to work with dual manifestations of some symmetry is well suited to 
deal with the chiral character of 2D anomalous gauge theories. Recently 
\cite{ABW} a new interpretation  for the phenomenon of dynamical mass 
generation known as Schwinger mechanism\cite{ls}, has been proposed which explores 
the ability of the soldering formalism to embrace interference effects.  
In that study the interference of right and left gauged Floreanini-Jackiw 
chiral bosons \cite{FJ} was shown to lead to a  massive vectorial mode, 
for the special case where the Jackiw-Rajaraman (JR) regularization parameter is
$a=1$ \cite{JR}\cite{RR}.

After the discovery that the $\chi QED_{2}$ could be consistently 
quantized if the regularization ambiguity were properly taken into account, the investigation 
on this subject has received considerable attention and emphasis \cite{many}\cite{AAR}.
The quantization of the model was considered from different points of view, 
both canonical and functional and the spectrum and unitarity was analyzed by distinct techniques, including the 
gauge invariant Wess-Zumino formulation\cite{WZ}, with results consistent with 
Ref.\cite{JR}.  Despite this spate of interest, a surprising new 
result was reported recently by Mitra \cite{PM} showing that a different 
regularization prescription was yet possible, leading to new consequences.   
He proposed a new (faddeevian) regularization class, materialized by a unique and
conveniently chosen mass term leading to a canonical description with three constraints.
Recall that in \cite{JR} and \cite{RR}, the Hamiltonian framework was structured in terms of
two classes with two ($a > 1$) and four ($a=1$) second-class constraints respectively. Mitra's work brings a 
clear interpretation for the reasons leading the bosonization ambiguity 
to fit into three instead of two distinct classes, classified according to the 
number of constraints present in the model.

It is the main goal of this paper to study the restrictions posed by the 
soldering formalism over this new regularization class.
Since soldering has ruled out the two-constraint class solution of Jackiw-Rajaraman as being able to dynamically generate mass via right-left interference, we are led to ask if the new Faddeevian class of chiral bosons proposed by Mitra do interfere constructively to produce a massive vectorial mode.  To find an answer to this question we review, in Section 2, 
the procedure of \cite{MM} to obtain the multi-parametric regularization effective action based on the Pauli-Villars regularization proposed in \cite{FS}. This effective action is the point of departure for an  
extention of the ambit of Ref.\cite{PM} that 
is needed to our purpose in this paper and to be developed in Section 3. The bosonised theory satisfying Faddeev's structure for the constraint algebra is studied in the canonical approach. The mass of the photon scalar field is computed and its dependence on the ambiguity parameter is shown to be tantamount to that in Ref.\cite{JR};  
the massless sector however is more constrained than its counterpart in \cite{JR},
corroborating the results of \cite{PM}.  The restrictions imposed by the 
soldering are worked out in Section 4.  We find the striking new result that the interference effects lift the parameter dependence by discriminating the value of the only non ambiguous class. Our results give a clear interpretation for the Schwinger mechanism as a left-right interference phenomenon, as suggested by Jackiw\cite{DTMP}. Our findings are further discussed in the final section.

\section{The Effective Action}

In a gauge invariant theory, free of anomalies, the canonical description reveals a couple of first-class constraints, with the Gauss law $G(x)$ appearing as the secondary constraint for the momentum $\pi_0(x)$ corresponding to the scalar potential $A_0(x)$.  In an anomalous gauge theory, on the contrary, gauge invariance is lost and the constraint algebra for the gauge generator becomes afflicted by the presence of a Schwinger term

\ba
\label{G1}
\left [G(x) , \pi_0(y)\right ] &=& 0\nonumber\\
\left [ G(x) , G(y)\right ] &=& \imath \, \hbar \,{\cal C} \,\delta '(x-y) ,
\ea
where ${\cal C}$ is some constant. This structure introduces extra degrees of freedom into the quantum theory as argued by Faddeev\cite{LF}.  The quantum chiral Schwinger model with the usual regularization ($a \geq 1$) does have more degrees of freedom than its classical counterparts, as expected, but does not fit into Faddeev's scheme above due to the functional dependence of the Gauss generator on the scalar potential, which leads to a different constraint algebra than (\ref{G1}),

\ba
\label{G2}
\left [G(x) , \pi_0(y)\right ] &\not=& 0\nonumber\\
\left [ G(x) , G(y)\right ] &=& 0 .
\ea
The second-class nature of the set is then due to the non-commutative character of the primary and secondary constraints. 

The new regularization class for the fermionic determinant proposed by Mitra has the virtue of fitting perfectly into Faddeev's picture.  In this section we shall review the computation of the fermionic determinant leading to this new scheme.
Our starting point is the action for fermionic sector of the chiral Schwinger model,
\be
S = \int\,d^{2}x\,\,\bar{\psi}(x)\,\left[
i\dslash - q\,\sqrt{\pi}\,\aslash(x)\,\left(1 + i\,\gamma_{5}\right)\right]\,
\psi(x)
\label{2.1}
\ee
where $\psi(x)$ is a fermionic field and $A_{\mu}$ is the vector gauge field
in a (1 + 1) dimensional spacetime. From this classical action
we obtain the following effective action\cite{FS}
\be
\exp\,i\,S_{eff}^{(0)}[A(x)] =
\int\,{\cal D}\psi(x)\,{\cal D}{\bar\psi}(x)\,
\exp\,i\,S\left[{\bar\psi}(x),\psi(x),A(x)\right].
\label{2.2}
\ee
In a formal level this is a nonlocal action that reads,
\be
\exp\,i\,S_{eff}^{(0)}[A(x)] = - q^{2}\,\int\,d^{2}x\,A^{\mu}(x)\,
\left(\eta_{\mu\alpha} + \epsilon_{\mu\alpha}\right)\,
\frac{\partial^{\alpha}\partial^{\beta}}{\partial^{2}}\,
\left(\eta_{\beta\nu} - \epsilon_{\beta\nu}\right)\,
A^{\nu}(x) ,
\label{2.3}
\ee
but there is an ambiguity related to the regularization procedure adopted.  Let us discuss the regularization procedure 
proposed by Frolov and Slavnov\cite{FS}. To this end we add a multi-parametric
regularising action
\be
S_{reg}[A(x)] = \sum_{r=1}^{2n-1}\int\,d^{2}x\,\,
{\bar \psi}_{r}(x)\,\left[ i\,\dslash - m_{r} - q\,\sqrt{\pi}\,
A_\mu(x)\,\Gamma_{r}^{\mu}\right]\,\psi_{r}(x)
\label{2.4}
\ee
where
\be
\Gamma_{r}^{\mu} = \left[a_{r}\,K^{\mu\nu}\,
\left(1 + i\gamma_{5}\right) + b_{r}\,\Sigma^{\mu\nu}\,
\left(1 - i\gamma_{5}\right)\right]\,\gamma_{\nu} .
\label{2.5}
\ee
Here $\psi_{r}(x)$ are the regulators fields with mass $m_{r}$ whose couplings
$\Gamma_{r}^{\mu}$ (or $K^{\mu\nu}$ and $\Sigma^{\mu\nu}$) are matrices
which will be determined later. These regulators bring up the following 
partition function
\be
\exp\,i\,S_{reg}^{eff}[A] =
\int\,\Pi_{r}\,{\cal D}\psi(x)\,{\cal D}{\bar\psi}(x)\,
\exp\,i\,S_{reg}\left[{\bar\psi}_{r}(x),\psi(x)_{r},A(x)\right]
\label{2.6}
\ee
which can be solved to \cite{MM}
\be
S_{reg}^{eff}[A] = -q^{2}\,\frac{\pi}{2}\,\int\,d^{2}x\,
A_{\mu}(x)\,G^{\mu\nu}(x,y)\,A^{\nu}(y)
\label{2.7}
\ee
with
\be
G^{\mu\nu}(x,y) = \int\,\frac{d^{2}p}{(2\pi)^{2}}\,{\bar G}^{\mu\nu}(p)\,
\exp\left[-i \cdot p\, (x - y)\right] .
\label{2.8}
\ee
Now  ${\bar G}_{\mu\nu}(p)$ is found to be
\[
{\bar G}^{(r)}_{\mu\nu}(p) = \frac{1}{\pi}\,\left\{\left(
a_{r}^{2}\,T^{1}_{\mu\nu\lambda\kappa} + b_{r}^{2}\,T^{1}_{\mu\nu\lambda\kappa}
\right)\,\left[2\,\left(1 + A_{r}\right)\left(\eta^{\lambda\kappa} - 
\frac{p^{\lambda}\,p^{\kappa}}{p^{2}}\right) +
A_{r}\,\eta^{\lambda\kappa}\right]+ 2A_{r} a_{r} b_{r} M_{\mu\nu}\right\}
\]
where
\be
A_{r} = 1 - \frac{i}{y_{r}}\,\ln(-1) + {\cal O}(y_{r})
\ee
and also
\begin{eqnarray}
T_{\mu\nu\lambda\kappa}^{1} &=& K_{\mu\rho}\,
\left(\delta_{\lambda}^{\rho} + \epsilon_{\kappa}^{\sigma}\right)\,
K_{\nu\sigma}\,
\left(\delta_{\kappa}^{\sigma} + \epsilon_{\kappa}^{\sigma}\right) \nonumber \\
T_{\mu\nu\lambda\kappa}^{2} &=& \Sigma_{\mu\rho}\,
\left(\delta_{\lambda}^{\rho} + \epsilon_{\kappa}^{\sigma}\right)\,
\Sigma_{\nu\sigma}\,
\left(\delta_{\kappa}^{\sigma} + \epsilon_{\kappa}^{\sigma}\right) \nonumber \\
M_{\mu\nu} &=& \left[
K_{\mu\lambda}\,\left(\eta^{\lambda\kappa} - \epsilon^{\lambda\kappa}\right)\,
\Sigma_{\nu\kappa} + \Sigma_{\mu\lambda}\,
\left(\eta^{\lambda\kappa} - \epsilon^{\lambda\kappa}\right)\,
K_{\nu\kappa}\right] \nonumber \\
y_{r}^{2} &=& \frac{p^{2}}{m_{r}^{2}} .
\end{eqnarray}
Imposing the conditions \cite{FS}
\ba
\sum_r \epsilon_r a_r^2 &=& \sum_r \epsilon_r b_r^2=0\nonumber\\
\sum_r \epsilon_r m_r a_r^2 &=& \sum_r \epsilon_r m_r b_r^2= \sum_r \epsilon_r m_r a_r b_r =0\nonumber\\
2\sum_r \epsilon_r a_r b_r &=& 1
\ea
where $\epsilon_r = (-1)^{r+1}$ is the Grassman parity and
then letting $m_{r} \rightarrow \infty$ we get,
\be
S_{reg}^{eff}[A] = q^{2}\,\frac{1}{2}\,\int\,d^{2}x\,A_{\mu}(x)\,
M^{\mu\nu}(x,y)\,A^{\nu}(y).
\label{2.16}
\ee
Jackiw and Rajaraman found a regularized solution with a 
diagonal choice for the matrix
\ba
M^{\mu\nu}\,=\,\pmatrix{a & 0 \cr
         		0 & a \cr}\, \delta(x - y),
\label{Matrix0}
\ea
with $a\geq 0$, corresponding to the cases with two and four-constraint's classes.  The physical content of these cases, as disclosed by them, was found to correspond to an $a$-dependent massive photon field and a massless fermion for the former, while in the later the photon field was absent. Mitra noticed that the alternative choice
\ba
M^{\mu\nu}\,=\,\pmatrix{1 & -1 \cr
         		-1 & -3 \cr}\, \delta(x - y),
\label{Matrix1}
\ea
leads to a new class of solutions with three second-class constraints 
and found that the physical spectrum of the model contains a chiral 
fermion and a photon field with mass $m=4\,q^2$. To work out the soldering 
formalism and obtain the interference contribution coming from the chiral 
fermions we need to generalize the regularization dependence of the effective 
action.  This is done in the next section.

\section{Hamiltonian Analysis and Spectrum}

In their seminal work Jackiw and Rajaraman\cite{JR} showed that the $\chi QED_{2}$ could be 
consistently quantized by including the bosonization ambiguity parameter satisfying the condition $a \geq 1$ to avoid tachyonic 
excitations. Later on, working out the canonical structure of the model, 
Rajaraman\cite{RR} showed that the cases $a>1$ and $a=1$ belonged to distinct classes: the $a=1$ case represents the four-constraints class, while the $a>1$ class presents only two constraints.  The latter is a continuous one-parameter class, while the former class is non ambiguous containing only one representative. The consequences 
of these distinct constraint structures are that the $a>1$ class presents, 
besides the massless excitation also a massive scalar excitation 
($m^2=\frac{e^2a^2}{a-1}$) that is not found on the other case.  
In the canonical approach the 
commutator between the primary and the secondary constraints vanishes in the 
first case.  The emergence of two more constraints 
completes the second-class set.  Mitra found the amazing fact that with an 
appropriated choice of the regularization mass term it is possible to 
close the second-class algebra with only three constraints.
His model is not manifestly Lorentz invariant, but the Poincar\'e generators have been constructed \cite{PM} and shown to close the relativistic algebra on-shell.
The main feature of this new regularization is the presence of a Schwinger term in the Poisson bracket algebra of the Gauss law, which limits the set to only three second-class constraints.  To see this let us 
write the CSM Lagrangian, with faddeevian regularization but with Mitra's 
regulator properly generalized to meet our purposes,
\be
{\cal L} = -\frac 14 \, F_{\mu\nu}\,F^{\mu\nu} + 
            \frac{1}{2}\,\partial_\mu\phi\,\partial^{\mu}\phi +
            q\,\left(g^{\mu\nu} + b\,\epsilon^{\mu\nu}\right)\,
            \partial_{\mu}\phi\,A_{\nu} +
           \frac{1}{2}\,q^2\,A_{\mu} M^{\mu\nu} A_{\nu}\,,
\label{Lagrangian}
\ee
where $F_{\mu\nu} = \partial_{\mu}A_{\nu}-\partial_{\nu}A_{\mu}$; $g_{\mu\nu} =
\mbox{diag}(+1,-1)$ and $\epsilon^{01} = -\epsilon^{10} = \epsilon_{10} = 1$.
$b$ is a chirality parameter, which can assume the values $b=\pm 1$. The
mass-term matrix $M_{\mu\nu}$ is defined as

\ba
M^{\mu\nu}\,=\,\pmatrix{1 & \alpha \cr
         		\alpha & \beta \cr}\, \delta(x - y).
\label{Matrix}
\ea
Notice that we have chosen unity coefficient for $A_{0}^{2}$ term. In a sense, this 
choice resembles Rajaraman's $a=1$ class and is the trademark of the
faddeevian regularization. In fact, Rajaraman's class is a singular point
in the ``space of parameters''. Its canonical description has the maximum
number of constraints with no massive excitation.
Such a case is found in (\ref{Lagrangian}) if we make $\alpha = 0$ in (\ref{Matrix}). The 
appearance of a new class needs a non vanishing value for $\alpha$.
With Mitra's choice, $\alpha = -1$ and $\beta = -3$, the photon becomes  massive ($m^2=4 \, q^2$),
but the remaining massless fermion has a 
definite chirality, opposite to that entering the interaction with the 
electromagnetic field. This choice is, however, too restrictive and may be relaxed 
leading to new and interesting consequences.  In this work the coefficients
$\alpha$ and $\beta$ are in principle arbitrary, but 
the mass spectrum will impose a  constraint 
between them.  This is best seen in the Hamiltonian formalism.

The canonical Hamiltonian is readily computed
\begin{eqnarray}
H = \int\!\!\! &dx& \!\!\! \left\{
\frac{1}{2}\,\left(\pi^{1}\right)^{2} + 
\frac{1}{2}\,\pi_{\phi}^{2} +
\pi^{1}A_{0}^{'} + 
\frac{1}{2}\,\phi^{'2} + q\,(\,b\phi'\,-\,\pi_{\phi}\,)\,A_0 +
q\left(\phi^{'} - b\,\pi_{\phi}\right)A_{1} 
+ \right. \nonumber \\
& & \left. +\,q^{2}\left(b - \alpha\right)A_{0}A_{1} +
\frac{1}{2}q^{2}\left(1-\beta\right)A_{1}^{2}
\right\}\, .
\label{Hamiltonian}
\end{eqnarray}
The stationarity algorithm leads a set of three constraints 
\begin{eqnarray}
\label{omega}
\Omega_{1} &=& \pi^{0}  , \nonumber \\
\Omega_{2} &=& \left(\pi^{1}\right)^{'} + 
               q\left(\pi_{\phi}-\,b\,\phi^{'}\right) -
               q^{2}\left(b - \alpha\right)A_{1}  , \\
\Omega_{3} &=&  -\left(b -\alpha\right)\,\pi^{1} + 
               2\,\alpha\,A_{0}^{'} + 
               \left(1 +\beta\right)A_{1}^{'}  ,\nonumber 
\end{eqnarray}
which are easily seen to be second-class, viz.
\begin{eqnarray}
\left\{\Omega_{1}(x), \Omega_{3}(y)\right\} &=& 2\,\alpha\,
\frac{\partial}{\partial x}\,\delta (x-y)\, ,  \nonumber \\  
\left\{\Omega_{2}(x), \Omega_{2}(y)\right\} &=& -2\,q^2\,\alpha\,
\frac{\partial}{\partial x}\,\delta (x-y)\, , \nonumber \\
\left\{\Omega_{2}(x), \Omega_{3}(y)\right\} &=& 
q^{2}\left(b-\alpha\right)^{2}\,\delta (x-y)-
\left(1+\beta\right)\,\frac{\partial^{2}}{\partial x \, \partial y}\,
\delta (x-y)\, , \nonumber \\
\left\{\Omega_{3}(x), \Omega_{3}(y)\right\} &=& -2\left(b-\alpha\right)
\left(1+\beta\right)\frac{\partial}{\partial x}\,\delta (x-y)\,\,,
\end{eqnarray}
with the other brackets vanishing. This is in sharp contrast with the usual regularization possessing two or four second-class constraints.  To perform quantization we compute the Dirac 
brackets,
\begin{eqnarray}
\label{db}
\left\{\,\phi(x)\,,\,\phi(y)\,\right\}_{D} &=& -\frac{1}{4\,\alpha}\,
                                       \theta (x-y)\, , \nonumber \\ 
\left\{\phi(x),A_{1}(y)\right\}_{D} &=& -\frac{1}{2\,q\,\alpha}\,
                                       \delta (x-y)\, , \nonumber \\ 
\left\{\phi(x),\pi^{1}(y)\right\}_{D} &=& -\frac{q}{4\,\alpha}
                                       \left(b - \alpha\right)\,
				       \theta (x-y)\, , \nonumber \\ 
\left\{A_{1}(x),A_{1}(y)\right\}_{D} &=& \frac{1}{2\,q^{2}\,\alpha}\, 
                                        \frac{\partial}{\partial x}\,
				         \delta (x-y)\, , \\  
\left\{\pi^{1}(x),A_{1}(y)\right\}_{D} &=& -\left(\frac{b+\alpha}{2\,\alpha}
                                           \right)\delta (x-y)\, , 
\nonumber \\ 
\left\{\pi^{1}(x),\pi^{1}(y)\right\}_{D} &=& -\frac{q^{2}}{4\,\alpha}
                                         \left(b-\alpha\right)^{2}
                                         \theta (x-y)\,\,. \nonumber   
\end{eqnarray}

The reduced Hamiltonian is obtained by strongly eliminating $\pi^{0}$, $A_{0}^{'}$
and $\pi_{\phi}$ from the constraints (\ref{omega}) and substituting in the canonical
Hamiltonian (\ref{Hamiltonian}),
\begin{eqnarray}
\label{hr}
H_{r} = \int\!\!\! &dx& \!\!\! \left\{
\frac{1}{2}\left(\pi^{1}\right)^{2} - 
\alpha\,\pi^{1}A_{1}^{'} + q\left(1-b\,\alpha\right)A_1\phi^{'} + 
\phi^{'2}-\frac{b}{q}\,\phi^{'}
\left(\pi^{1}\right)^{'} +
 \right. \nonumber \\
&& \left. + \frac{1}{2\,q^{2}}
\left(\pi^{1}\right)^{'2} +\frac{1}{2}\,q^{2}\left(\alpha^{2}-
\beta\right)A_{1}^{2}\right\}\, .
\end{eqnarray}
Making use of (\ref{db}) and (\ref{hr}) we get the following equations of
motion for the remaining fields,
\begin{eqnarray}
\dot{\phi} &=& b\,\phi^{'} - \frac{1}{q} 
\left(\pi^{1}\right)^{'} + \frac{q}{2\,\alpha}
\left(1 - 2\alpha^{2}+\beta\right)A_{1} \label{phi_eq}\, ,\label{phi} \\
\dot{\pi}^{1} &=& -b
\left(\pi^{1}\right)^{'} +\frac{q^{2}}{2\,\alpha}\left[(b-\alpha)
(1-\alpha^{2}) - (b+\alpha)(\alpha^{2}-\beta)\right]A_{1} \label{pi1}\, , \\
\dot{A}_{1} &=& \left(\frac{\alpha + b}{2\alpha}\right)\pi^{1} -
\left(\frac{1 + \beta}{2\alpha}\right)A_{1}^{'}\,\, . \label{a1}
\end{eqnarray}

We are now ready to determine the spectrum of the model. Isolating $\pi^{1}$
from the Eq.(\ref{a1}) and substituting 
in the equation (\ref{pi1}), we will have
\begin{eqnarray}
\left(\frac{2\,\alpha}{\alpha + b}\right)\ddot{A}_{1}+
b\left(\frac{1 + \beta}{\alpha + b}\right)A_{1}^{''}\!\!\! &=& \!\!\!
-\left(\frac{2\,b\,\alpha}{\alpha + b} + \frac{1+\beta}{\alpha +b}\right)
\dot{A}_{1}^{'}\,+\, \nonumber \\
\!\!\! & + & \!\!\! \frac{q^2}{2\alpha}\left[\left(b - \alpha\right)
\left(1-\alpha^{2}\right) - \left(b + \alpha\right)
\left(\alpha^{2}-\beta\right)\right]A_{1}\, . 
\label{eq}
\end{eqnarray}
To get a massive Klein-Gordon equation for the photon field
we must set 
\be
\label{condicao}
\left(1 + \beta\right) + b\,\left(2\alpha\right) = 0\, ,
\ee
which relates $\alpha$ and $\beta$ and shows that the regularization ambiguity adopted in 
\cite{PM} can be extended to a continuous one-parameter class (for a chosen 
chirality). We have, using (\ref{eq}) and (\ref{condicao}), the following mass 
formula for the massive excitation of the spectrum,
\be
m^{2} = q^{2}\,\frac{\left(1 + b\,\alpha\right)^{2}}{b\,\alpha}\, .
\ee
Note that to avoid tachyonic excitations, $\alpha$ is further restricted to 
satisfy $b\,\alpha = |\alpha|$, so $\alpha \rightarrow -\alpha$ 
interchanges from one chirality to another.
Observe that in the limit $\alpha \rightarrow 0$ the massive 
excitation becomes infinitely heavy and decouples from the spectrum.  This 
leads us back to the four-constraints class. It is interesting to see that the redefinition of 
the parameter as $a=1+ |\alpha|$ leads to,
\be
m^2 = \frac{q^2 a^2}{a-1}
\ee
which is the celebrate mass formula of the chiral Schwinger model, showing that the parameter dependence of the mass spectrum 
is tantamount to both the Jackiw-Rajaraman and the faddeevian 
regularizations.

Let us next discuss the massless sector of the spectrum. To disclose the 
presence of the chiral excitation we need to diagonalize the  reduced 
Hamiltonian (\ref{hr}).  This procedure may, at least in principle, impose further restrictions over $\alpha$.
This all boils down to find the correct  linear 
combination of the fields leading to the free chiral equation of motion.  To this end 
we substitute $\pi^{1}$ from its definition and $A_{1}$ from
the Klein-Gordon equation into equation (\ref{phi_eq}) to obtain
\begin{eqnarray}
\label{vinte}
0 &=&\frac{\partial}{\partial t}\left\{
\phi  +  \frac{q}{2\alpha}
\left(\frac{2 + 2\,b\,\alpha - \alpha^{2}}{m^{2}}\right)
\dot{A}_{1} + \frac{1}{q}\left(\frac{\alpha}{\alpha + b}
\right)A_{1}^{'}\right\}\,-\,  \nonumber \\
&-& \frac{\partial}{\partial x}\left\{
b\,\phi - \frac{1}{q}\left(\frac{\alpha}{\alpha + b}\right)\dot{A}_{1} +
\left[\frac{q}{2\alpha}\left(\frac{2 + 2\,b\,\alpha -\alpha^{2}}{m^{2}}\right)
-\frac{1}{q}\left(\frac{2\,b\,\alpha}{\alpha+b}\right)\right]\right\}\,\,.
\end{eqnarray}
This expression becomes the equation of motion for a self-dual boson $\chi$ 
\be
\dot{\chi} - b\,\chi^{'} = 0
\label{quiral}
\ee
if we identify the coefficients for $\dot A_1$ and $A_1^{'}$ in the two independent terms of (\ref{vinte}) with,

\be
\label{chi}
\chi = \phi  +  \frac 1{q} \left(\frac{\alpha}{\alpha + b}\right)\left(A_{1}^{'}- b \dot{A}_{1}\right).
\ee
This field redefinition, differently from the case of the massive field whose construction imposed condition (\ref{condicao}), does not restrain the parameter $\alpha$ any further.
Using 
(\ref{omega}) and (\ref{quiral}), all the fields can be expressed as 
functions of the free massive scalar $A_1$ and the free chiral boson 
$\chi$, interpreted as the bosonized Weyl fermion.  The main result of this section is now complete, i.e., the construction of the one-parameter class regularization generalizing Mitra's proposal. The stage is now set to study the interference of chiral actions with (one-parameter) faddeevian regularization.

\section{Effects of Interference}

In this section we use the soldering formalism introduced in
\cite{ms} to examine the restriction imposed by chiral interference over
the regularization ambiguity parameter when the faddeevian approach is adopted.
This study, taken in the framework of the usual JR regularization, establishes a strong restriction over the parameter's values and gives rise to a new interpretation
for the mechanism of dynamical mass generation occurring in the Schwinger model.
This study is meaningful and necessary since a new class of theories with three
second-class constraints has emerged: it must be verified if new solutions resulting from interference will lead to a gauge invariant massive excitation. To begin with, let
us rewrite explicitly the two chiral actions presented in (\ref{Lagrangian})
in the appropriate light-cone variables,

\ba
{\cal L}_+\,&=& \partial_{+}\rho\,\partial_{-}\rho + \frac{1}{2}\,
\left(\partial_{-}A_{+} - \partial_{+}A_{-}\right)^{2} +
2\,q\,\partial_{-}\rho\,A_{+} 
-2\, q^{2}\,|\alpha|\,A_{-}^{2} +\nonumber \\ 
& &+\;q^{2}\,\left(1\,+\,|\alpha|\right)A_{-}A_{+} \label{mitra+}\\
{\cal L}_-\,&=&\partial_{+}\varphi\,\partial_{-}\varphi + \frac{1}{2}\,
\left(\partial_{-}A_{+} - \partial_{+}A_{-}\right)^{2} +
2\,q\,\partial_{+}\varphi\,A_{-} 
-2\, q^{2}\,|\bar\alpha|\,A_{+}^{2} +\nonumber \\
& &+\;q^{2}\,\left(1\,+\,|\bar\alpha|\right)A_{-}A_{+} \;\;,
\label{mitra-}
\ea
where we have used the convention ${\cal L}_{\pm} = {\cal L}|_{b}$. For clarity, we have used different fields $(\varphi, \rho)$ for opposite chiralities and the corresponding 
mass-term parameters $(\alpha,\overline{\alpha})$ to make clear that these
chiral theories are uncorrelated. However by making use of soldering
formalism we will get a meaningful combination of these components.

The main point of soldering is to lift the global Nother symmetry of each chiral component to a local symmetry of the system as a whole. Showing only the main parts of the soldering 
formalism we can see that the axial transformation ($\delta\varphi =\delta\rho = \eta$) leads to

\ba
\label{CW10}
\delta {\cal L}_+\,&=&\, \partial_-\eta\,J_+(\rho) \nonumber \\
\delta {\cal L}_-\,&=&\, \partial_+\eta\,J_-(\varphi)
\ea
where $J_{-}(\varphi)=2(\partial_-\varphi\,+\,q A_-)$ and $J_{+}(\rho)=2(\partial_+\rho\,+\,q A_+)$ are the Noether's currents and $\eta$ is the gauge parameter. Next we introduce the soldering field
$B_{\pm}$ appropriately coupled to the Noether currents to obtain the once iterated chiral actions as,

\ba
{\cal L}_+^{(0)} &\rightarrow& {\cal L}_+^{(1)}\,=\,{\cal L}_+^{(0)}\,+\,B_+\,
J_-(\varphi) \nonumber \\
{\cal L}_-^{(0)} &\rightarrow& {\cal L}_-^{(1)}\,=\,{\cal L}_-^{(0)}\,+\,B_-\,
J_+(\rho) \;\; .
\ea
The soldering fields act as partial compensators for the variance (\ref{CW10}), transforming vectorially under the axial symmetry, $\delta B_{\pm}=\partial_{\pm}\eta$.
It is now possible to define an effective Lagrangian
invariant under the combined transformation of the chiral fields and compensators as,

\be
{\cal L}_{eff}\,=\,{\cal L}_+^1\,+\,{\cal L}_-^1\,+\,2\,B_+\,B_-\;\;.
\label{Leff}
\ee
The soldered action is obtained using the fact that $B_{\pm}$ are auxiliary fields. Their elimination may be done altogether from their field equations but the effects of 
soldering will persist as a residual symmetry for the remaining fields. This will naturally cohere the otherwise
independent chiral fields $\varphi$ and $\rho$ in the form of a
soldered Lagrangian for a collective field $\Phi$ as,
\ba
\label{solda}
{\cal L}_{eff}\,&=&\,\partial_+\,\Phi\,\partial_-\,\Phi\,-\,2\,q\,
(A_+\,\partial_-\,
\Phi\,-\,A_-\,\partial_+\,\Phi)\,+\, \frac 12 \left(\partial_+ A_-\, -\, \partial_- A_+\right)^2\nonumber\\
\,&+&\,{q^2 \over 2}\left[\alpha\,A_+^2\, - \,\overline{\alpha}\,A_-^2\, - \,\left(\alpha - \,\overline{\alpha}\right) A_+\,A_-\,\right]
\ea
where $\Phi=\varphi-\rho$. Notice that except for the last term, the soldered action describes the massive gauge invariant bosonised version of the Schwinger model, with the gauge invariant collective field $\Phi$ playing the role the of the photon field. Gauge invariance imposes a strong constraint over the parameters as,
\be
\alpha\,= \,\overline{\alpha} = 0. 
\ee
This value corresponds to the $a=1$, four-constraints regularization class. This is a remarkable result, consistent with \cite{ABW}.

A notable feature of the present analysis is the disclosure of a new class of
parameterizations and their dependence with the number of constraints.
Different aspects of this feature were
elaborated and the consequences of interference computed.  
To discuss further the implications of interference on chiral actions it is best to compare with the existing literature. This also serves
to put the present work in a proper perspective.
To be precise, it was initialy shown that in the faddeevian approach there are actually three
second-class constraints with a real parameter dependence.
To disclose this one-parameter dependence of the faddeevian regularization is a new interesting result. The counting of constraints explains the presence of only one chiral excitation in the spectrum (besides the massive mode).
This is in contrast with the usual JR regularization where the massless excitation is scalar, and is essentially tied to the fact that this regularization is less constrained.

The restrictions of soldering however confine the appearance of a massive vector excitation to the interference of modes belonging to the $a=1$ class that, being more constrained, has only a massless scalar in the spectrum.
This might raise questions about the interference of the chiral modes in this class.
It should be noticed however that the use of light-cone variables in the soldering constrains even further these chiral actions.  Both the two and the four-constraints classes display chiral excitations instead of massless scalars.  The original chiral mode of the Mitra's class therefore disappear in the presence of the extra light-cone constraint and there appears to exist an ambiguity challenging the real meaning of the soldering.  In fact there are no massless particles in the spectrum 
of (\ref{mitra+}-\ref{mitra-}) for the light-cone setting.
However, what is important to observe in this scenario is that the whole process of soldering is done in the Lagrangian framework, such that the limit $a\rightarrow 1$ is well defined.  This is also valid for the JR regularization.  The limit leads to the $a=1$ action and the canonical analysis may be done unambiguously.  Oppositely, the Hamiltonian formulation has the $a=1$ point as a singularity, as shown in (\ref{db}).

\section{Conclusions}

In this work we studied the bosonized form of the CSM fermionic determinant adopting the three-constraints regularization parameterized by a single real number.  This extends early regularizations 
proposed by JR and Mitra.  Our results display a clear-cut separation of the 
existing classes shown to depend only on the number of 
second-class constraints. The new class with faddeevian regularization and three second-class constraints has been worked out in great detail.  The spectrum 
has been shown to consist of a chiral boson and a massive photon field.
The mass formula for the scalar excitation was shown to reproduce the JR result. Considerations of unitarity therefore restrain the range of the regularization parameter similarly.  The  use of the
soldering formalism supplemented by gauge invariance restricts the otherwise arbitrary ambiguity parameter to the specific value $a=1$, which corresponds to the four-constraints class. This is new result that discriminates the special character of this unambiguous regularization point and gives a precise interpretation of the Schwinger dynamical mass generation mechanism as a consequence of right and left interference.

To conclude we stress that the formalism and analysis proposed here illuminates the close
connection among anomalous gauge theories, the interference phenomenon and the mechanism
of dynamical mass generation, providing a variety of new possibilities with practical
applications.

\bigskip

\noindent {\bf Acknowledgment:}   This work is supported in part by
CNPq, FINEP, CAPES, FAPESP and FUJB (Brazilian Research Agencies).

\end{document}